\documentclass{optica-article}

\journal{opticajournal}

\articletype{Research Article}
\usepackage{lineno}
\usepackage[normalem]{ulem}

\begin{document}

\title{Large-scale Global Low-rank Optimization for Computational Compressed Imaging}

\author{Daoyu Li,\authormark{1,2} Hanwen Xu,\authormark{1,2} Miao Cao,\authormark{3} Xin Yuan,\authormark{3} David J. Brady,\authormark{4} and  Liheng Bian\authormark{1,2,*}}

\address{\authormark{1}School of Information and Electronics \& Advanced Research Institute of Multidisciplinary Sciences, Beijing Institute of Technology, 100081, Beijing, China\\
\authormark{2}MIIT Key Laboratory of Complex-field Intelligent Sensing, Beijing Institute of Technology, Beijing 100081, China\\
\authormark{3}Research Center for Industries of the Future and School of Engineering, Westlake University, 310024, Hangzhou, China\\
\authormark{4}Wyant College of Optical Sciences, University of Arizona, AZ 85721, Tucson, USA}

\email{\authormark{*}bian@bit.edu.cn} 



\begin{abstract}
Computational reconstruction plays a vital role in computer vision and computational photography. Most of the conventional optimization and deep learning techniques explore local information for reconstruction. Recently, nonlocal low-rank (NLR) reconstruction has achieved remarkable success in improving accuracy and generalization. However, the computational cost has inhibited NLR from seeking global structural similarity, which consequentially keeps it trapped in the tradeoff between accuracy and efficiency and prevents it from high-dimensional large-scale tasks. To address this challenge, we report here the \emph{global low-rank} (GLR) optimization technique, realizing highly-efficient large-scale reconstruction with global self-similarity. Inspired by the self-attention mechanism in deep learning, GLR extracts exemplar image patches by feature detection instead of conventional uniform selection. This directly produces key patches using structural features to avoid burdensome computational redundancy. Further, it performs patch matching across the entire image via neural-based convolution, which produces the global similarity heat map in parallel, rather than conventional sequential block-wise matching. As such, GLR improves patch grouping efficiency by more than one order of magnitude. We experimentally demonstrate GLR’s effectiveness on temporal, frequency, and spectral dimensions, including different computational imaging modalities of compressive temporal imaging, magnetic resonance imaging, and multispectral filter array demosaicing. This work presents the superiority of inherent fusion of deep learning strategies and iterative optimization, and breaks the persistent dilemma of the tradeoff between accuracy and efficiency for various large-scale reconstruction tasks.
\end{abstract}

\section{Introduction}
By jointly optimizing computation and optics, computational imaging \cite{altmann2018quantum} improves the efficiency and information capacity of optical systems \cite{wu2022integrated}. With computational imaging, one can achieve detection under extremely complex lighting conditions \cite{kirmani2014first,morris2015imaging4}, capture invisible high-dimensional information \cite{morris2015imaging4,zheng2013wide5,qiao2021evaluation6}, look through obstacles \cite{o2018confocal7,saunders2019computational8}, etc. As a critical tool for computational imaging, computational reconstruction recovers high-dimensional large-scale data from compressed or aliased low-dimensional signals. A typical reconstruction technique first formulates a physical model to describe how to obtain the measurement with a specific target scene based on geometrical or fluctuating optics. It derives target information of interest by solving the inverse problem of the forward imaging model \cite{mait2018computational9}. Despite a few cases that can be solved directly by inverse transformation, prior constraints are vital to regulate the reconstruction and eliminate noise and disturbance.

To date, there exist two kinds of most prevalent priors, including model-based and statistics-based ones. Conventional reconstruction techniques with model-based prior such as total variation (TV) \cite{rudin1992nonlinear10} and Tikhonov regularization \cite{Tikhonov1997NonlinearIP11}, mainly focus on local information of the image. Motivated by rich self-repetitive structures in natural images, exploiting the nonlocal self-similarity (NSS) model has led to the well-known nonlocal means (NLM) \cite{buades2005non12} and block matching and 3D filtering (BM3D) \cite{dabov2006image13,dabov2007image14}. The bloom of compressive sensing theories inspired the low-rank regularization \cite{gu2014weighted15,dong2014compressive16,zhang2018nonlocal17} of self-repetitive patches. Nonlocal low-rank (NLR) methods have shown state-of-the-art accuracy and strong generalization for various computational photography tasks \cite{yuan2020plug18,yuan2021snapshot19}. However, further extending the conventional nonlocal techniques to global perception results in unacceptable computational complexity especially in high-dimensional large-scale cases \cite{yuan2020plug18,yuan2021snapshot19,liu2018rank20}. With the bloom of deep learning, the most commonly used statistics-based approaches for computational reconstruction are learning-based methods \cite{barbastathis2019use21}. Training a neural network on a large-scale dataset has been proven to be an effective way to learn the inverse imaging model \cite{belthangady2019applications22,wang2020deep23}. However, there remains a challenge for learning-based approaches to construct a generalized network robust to different system settings and noise levels. Besides, concerns have been raised regarding the use of black-box deep learning for high-stakes tasks in healthcare, criminal justice, etc \cite{rudin2019stop}.

Bearing these concerns in mind, we present the \emph{global low-rank} (GLR) optimization for computational reconstruction. GLR regulates the optimization with global self-similarity prior, which incorporates the structural information of the entire image into reconstruction for each local image patch. GLR realizes efficient reconstruction with high accuracy and robustness, by attention from structural feature detection and similar patch matching with neural computation. These two operations promote running efficiency by up to above one magnitude of order for high-dimensional large-scale computational reconstruction with improved accuracy.

GLR first conducts structural feature attention by extracting exemplar edge patches according to corners of the edge maps (Fig. 1 \textbf{a}). Conventional NLR first uniformly picks exemplar patches as shown in Fig. 1 \textbf{a}. Mean square error (MSE) is utilized to measure the similarity between the exemplar patch and other patches. Corner-based exemplar patch selection is inspired by common sense that natural images have sparse edges and corners which locate in the area of significant textures. Corner-based exemplar patches contribute to focus on the recovery of sharp edges which are tougher than that of smooth areas. We experimentally demonstrate that nonlocal techniques with fewer corner-based exemplar patches can achieve competitive performance and reduced running time against more evenly arranged patches.

Next, as shown in Fig. \ref{fig:GM} \textbf{b}, GLR groups the patches across the entire image following structural similarity order. The block matching (BM) performs a k-nearest-neighbor (KNN) search for each exemplar patch within a local window\cite{dong2014compressive16}. Yuan et al.\cite{yuan2020plug18,yuan2021snapshot19} have experimentally proved that the nonlocal low-rank reconstruction based on vanilla BM for snapshot compressed imaging requires more than one hour to restore a video clip at 256$\times$256$\times$8 resolution. Benefiting from the bloom of deep learning framework, we can realize parallel patch matching with the ‘Conv2d’ layer (a neural layer performing batch convolution for 2-dimensional input), enabling fast and accurate computing. GLR improves the matching speed by an order of magnitude, with advanced accuracy to vanilla BM.

The simulation and experiment results demonstrate GLR exhibits boosted efficiency, high precision, and strong generalization for reconstruction, facilitating diverse computational photography applications. This work presents the superiority of inherent fusion of deep learning strategies and iterative optimization, and breaks the persistent dilemma of the trade-off between accuracy and efficiency for various large-scale reconstruction tasks.

\section{Method}
\begin{figure}[h!]
    \centering
    \includegraphics[width=1\linewidth]{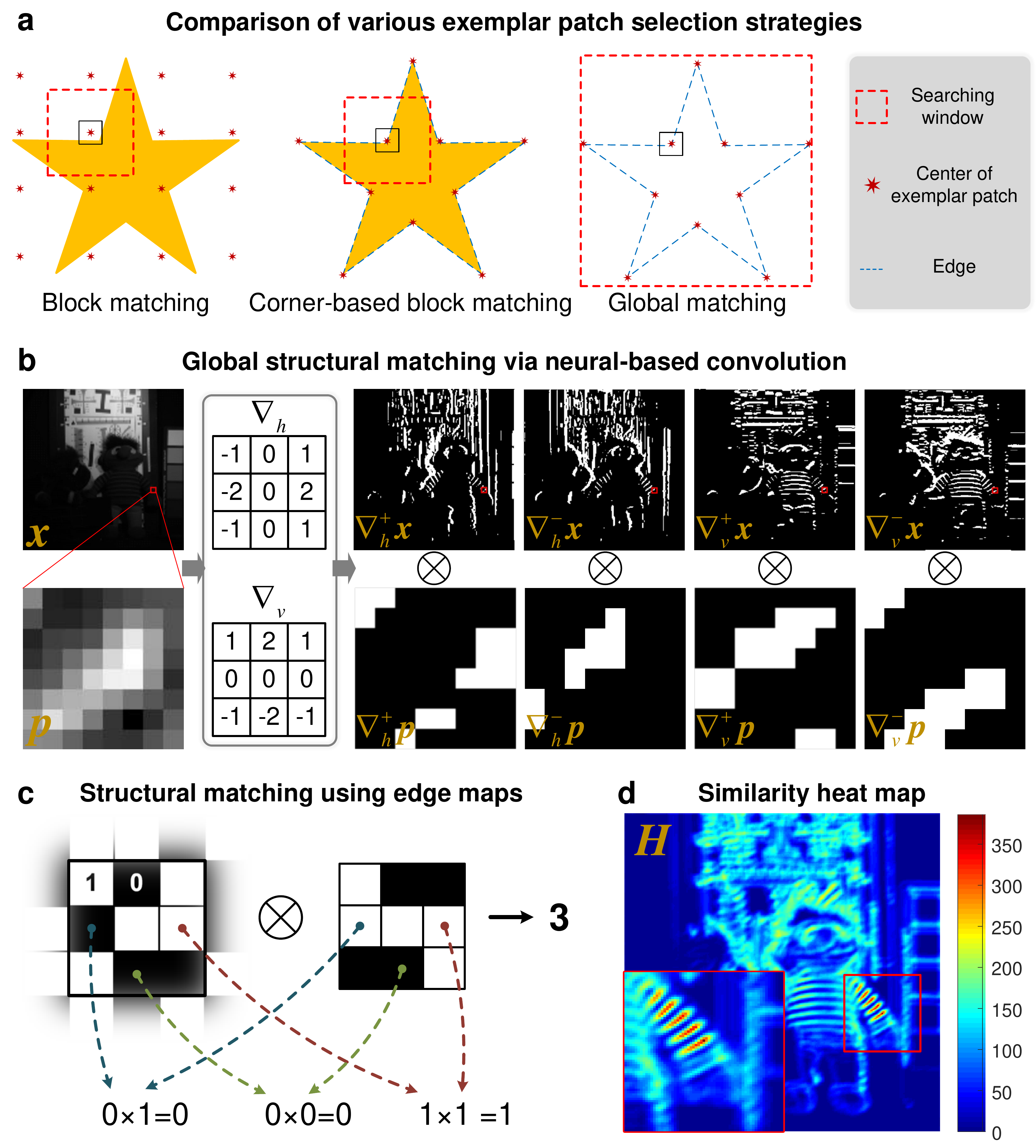}
    \caption{Illustration of global matching (GM). \textbf{a}, The comparison of various exemplar patch selection strategies. Vanilla block matching (BM) uniformly picks exemplar patches. Corner-based BM selects exemplar patches based on the corner points of the edge map. GM obtains the similarity heat maps by correlating the edge map of the image with exemplar edge patches. The search range is extended to the whole image. \textbf{b}, Both horizontal and vertical gradient edge maps ($\nabla_{h}\boldsymbol{x}$ and $\nabla_{v}\boldsymbol{x})$ and exemplar edge patches ($\nabla_{h}\boldsymbol{p}$ and $\nabla_{v}\boldsymbol{p}$) are extracted by the Sobel operators ($\nabla_h$ and $\nabla_v$). The similarity heat map for the exemplar patch is the summation heat maps corresponding to horizontal and vertical edge maps which are obtained by Conv2d operation $\bigotimes$. \textbf{c}, The exemplar edge patch slides over the edge map, performing an elementwise multiplication and obtaining the numbers of points of overlapped edges. \textbf{d}, The similarity heatmap of the exemplar patch in \textbf{a} extracted by GM.}
    \label{fig:GM}
\end{figure}

Recall that the conventional nonlocal technique performs a KNN search for each exemplar patch within a local window. It first uniformly picks exemplar patches as shown in Fig. \ref{fig:GM} \textbf{a}. The exemplar patch is the center patch of every local window. Mean square error (MSE) is utilized to measure the similarity between the exemplar patch and other patches.
However, it has relatively low efficiency when using low-value patches which do not contain textures and details as shown in Fig. \ref{fig:GM} \textbf{a}. For further improvement in the efficiency of matching, we introduce the exemplar patch selection based on corner points of the image’s edge map (Fig. \ref{fig:GM} \textbf{a}). Corner-based exemplar patch selection is inspired by common sense that natural images have sparse edges and corners which locate in the area of significant textures. Corner-based exemplar patches enable structural feature attention by focusing on the recovery of sharp edges which are tougher than that of smooth areas. We experimentally demonstrate that nonlocal techniques with fewer corner-based exemplar patches can achieve competitive performance and reduced running time against more evenly arranged patches in Section \ref{sec:results_sci}.

\begin{figure}[h]
	\centering
	\includegraphics[width=1\linewidth]{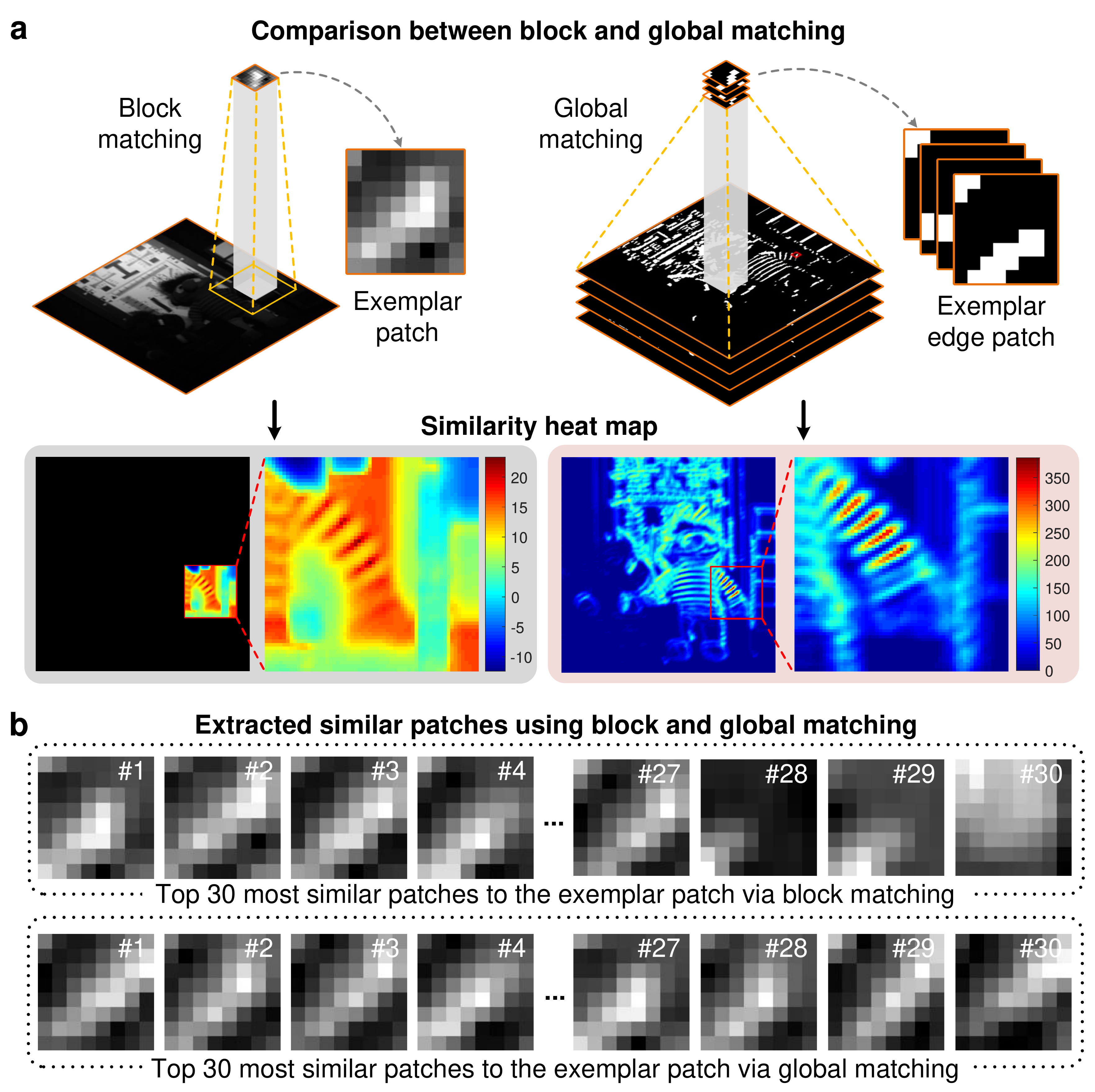}
	\caption{Comparison between BM and GM. \textbf{a}, BM searches similar patches by measuring the MSE of each patch against the exemplar patch within a local window. GM obtains the similarity heat maps by correlating the edge maps with exemplar edge patches. 
		The similarity heat map for the exemplar patch is the summation heat maps corresponding to horizontal and vertical edge maps which are obtained by neural-based Conv2d operation. \textbf{b}, Comparison between extracted similar patches from BM and the reported GM. The patches extracted by GM have higher structural similarity to the exemplar patch in \textbf{a}.}
	\label{fig:BMvsGM}
\end{figure}

The second limitation of nonlocal algorithms lies in the local receptive field of vanilla BM. Denote the $i$-th exemplar patch as $x_i$, and $x_j$ stands for the neighbor patches within a local window of $x_i$. 
\begin{equation}
	G_i = \left\lbrace i_j | \| x_i - x_j\|_2^2 < T_i\right\rbrace ,
\end{equation}
where $T_i$ is a pre-defined threshold, and $G_i$ is the collection of positions corresponding to those similar patches. Due to BM’s high computational complexity ($O(n^2)$), common practice sets the size of the local window to a relatively small value. To address this issue, GM efficiently search similar patches across the entire image. Given the to-be-updated image $\boldsymbol{x}$, we first extract the binary edge maps of the image ($\nabla_{h}\boldsymbol{x}$ and $\nabla_{v}\boldsymbol{x}$) by the Sobel operators ($\nabla_h$ and $\nabla_v$). 
\begin{equation}
		\begin{aligned}
			\nabla_{h}^{+}\boldsymbol{x} &= T\left( abs \left( \nabla_{h} \boldsymbol{x}\right)\right) , \\
			\nabla_{h}^{-}\boldsymbol{x} &= T\left( abs \left( - \nabla_{h} \boldsymbol{x}\right)\right) , \\
			\nabla_{v}^{+}\boldsymbol{x} &= T\left( abs \left( \nabla_{v} \boldsymbol{x}\right)\right) , \\
			\nabla_{v}^{-}\boldsymbol{x} &= T\left( abs \left( - \nabla_{v} \boldsymbol{x}\right)\right).
		\end{aligned}
            \label{eq:6}
\end{equation}
$T$ is the threshold function
	\begin{equation}
		T\left( z\right) = \left\{
		\begin{aligned}
			1, \ z>th, \\
			0, \ z\leq th,
		\end{aligned}
		\right.
            \label{eq:7}
	\end{equation}
 where $th$ is the pre-defined threshold. GM picks exemplar patches centered on the corner points of the averaged edge map rather than traditional uniform selection. After exemplar patches selection, we perform global patch matching using the convolution operation of deep learning frameworks. As shown in Fig. \ref{fig:BMvsGM} \textbf{a}, each exemplar edge patch ($\nabla_{h}\boldsymbol{p}$ and $\nabla_{v}\boldsymbol{p}$) acts as a convolution kernel to convolute the edge map. Note that the convolution operator has slightly different from the convolution in mathematics. It specifically refers to the ‘Conv2d’ in deep convolutional networks. Benefiting from the bloom of deep learning framework, we can perform parallel convolution by concatenating the exemplar edge patches $\nabla\boldsymbol{p}$ into a multi-channel convolutional kernel $\nabla\boldsymbol{P}$. Figure \ref{fig:BMvsGM} \textbf{b} demonstrates that GM enables increasing matching precision and finding more structurally similar patches than BM. In the case of multi-channel images, both the edge maps and the exemplar edge patches are of three dimensions. We can conduct the batch convolution with the multi-channel edge maps as batch inputs. The similarity heat map is obtained as
\begin{equation}
		\boldsymbol{H} = \sum_{a=\left( h,v\right), b=\left( +,-\right) } \text{Conv2d}\left( \nabla_{a}^{b}\boldsymbol{x}, \enspace \nabla_{a}^{b}\boldsymbol{P} \right). 
        \label{eq:8}
\end{equation}
The sizes of $\nabla_{a}^{b}\boldsymbol{x}$, $\nabla_{a}^{b}\boldsymbol{P}$, and $H$ are H$\times$W$\times$C, P$\times$P$\times$C$\times$N, and (H-P+1)$\times$(W-P+1)$\times$N respectively, where H$\times$W is the size of input images, P$\times$P is the size of extracted patches, C represents the input channels, and N denotes the number of exemplar patches. GM obtains all the heat maps of N exemplar patches with 4 Conv2d operations while BM requires N times KNN-based matching. So the neural-based convolution contributes to reduce computation load and boost the running efficiency of GLR.
Besides, modern deep learning frameworks provide several acceleration approaches to speed up the Conv2d layer including Im2Col + matrix manipulation\cite{vedaldi2015matconvnet51}, fast Fourier transform (FFT), Winograd\cite{lavin2016fast52}, etc. A larger value in the heat map means that the image patch centered at the corresponding position has a higher structural similarity to the exemplar patch. 

We use the operator $\widetilde{\boldsymbol{R}_{i}}$ to represent the $i$-th global matching operation. The resulted grouped matrix is $\widetilde{\boldsymbol{R}_{i}}\boldsymbol{x}$, where $\boldsymbol{x}$ denotes the to-be-update image. The objective function of low-rank approximation is 
\begin{equation}
	\boldsymbol{x} = \underset{\boldsymbol{x}}{\arg\min } \sum_{i} \operatorname{rank}\left(\widetilde{\boldsymbol{R}_{i}} \boldsymbol{x}\right),  \text {  s.t.  }  \boldsymbol{y}=\boldsymbol{\Phi} \boldsymbol{x},
	\label{eq:low-rank}
\end{equation}
where $\boldsymbol{y}$ is the measurement, and $\boldsymbol{\Phi}$ stands for the sensing matrix. We adopt WNNM \cite{gu2014weighted15} for solving the above low-rank approximation in Eq. (\ref{eq:low-rank}). The reconstructed image is derived by the iterative optimization framework such as generalized alternative projection (GAP) \cite{yuan2016generalized25} and alternating direction multiplier (ADMM) \cite{boyd2011distributed26}.


\section{Results}
We applied GLR and the existing CS-based algorithms on both simulation and experiment data of three computational imaging tasks including coded aperture compressive temporal imaging (CACTI)\cite{llull2013coded27}, magnetic resonance imaging (MRI)\cite{weiskopf2021quantitative28}, and multispectral filter array (MSFA) demosaicing\cite{lapray2014multispectral29}. In the following evaluations, we employed the peak signal-to-noise ratio (PSNR) and structural similarity index (SSIM) \cite{wang2004image42} to quantify reconstruction accuracy. In the evaluations of the Results and the following Discussion sections, GLR and NLR shared the same hyper-parameters. The deep learning platform MatConvNet\cite{vedaldi2015matconvnet51} is utilized for convolution-based GM. All the calculations are done on a desktop PC with an Intel i7-9700K CPU and 64G RAM for a fair comparison between GLR and conventional nonlocal low-rank techniques. We believe that GLR can be faster on convolution-accelerating hardware.

\subsection{Coded aperture compressive temporal imaging} \label{sec:results_sci}
\begin{figure*}[h!]
    \centering
    \renewcommand{\floatpagefraction}{.9}
    \includegraphics[width=1\linewidth]{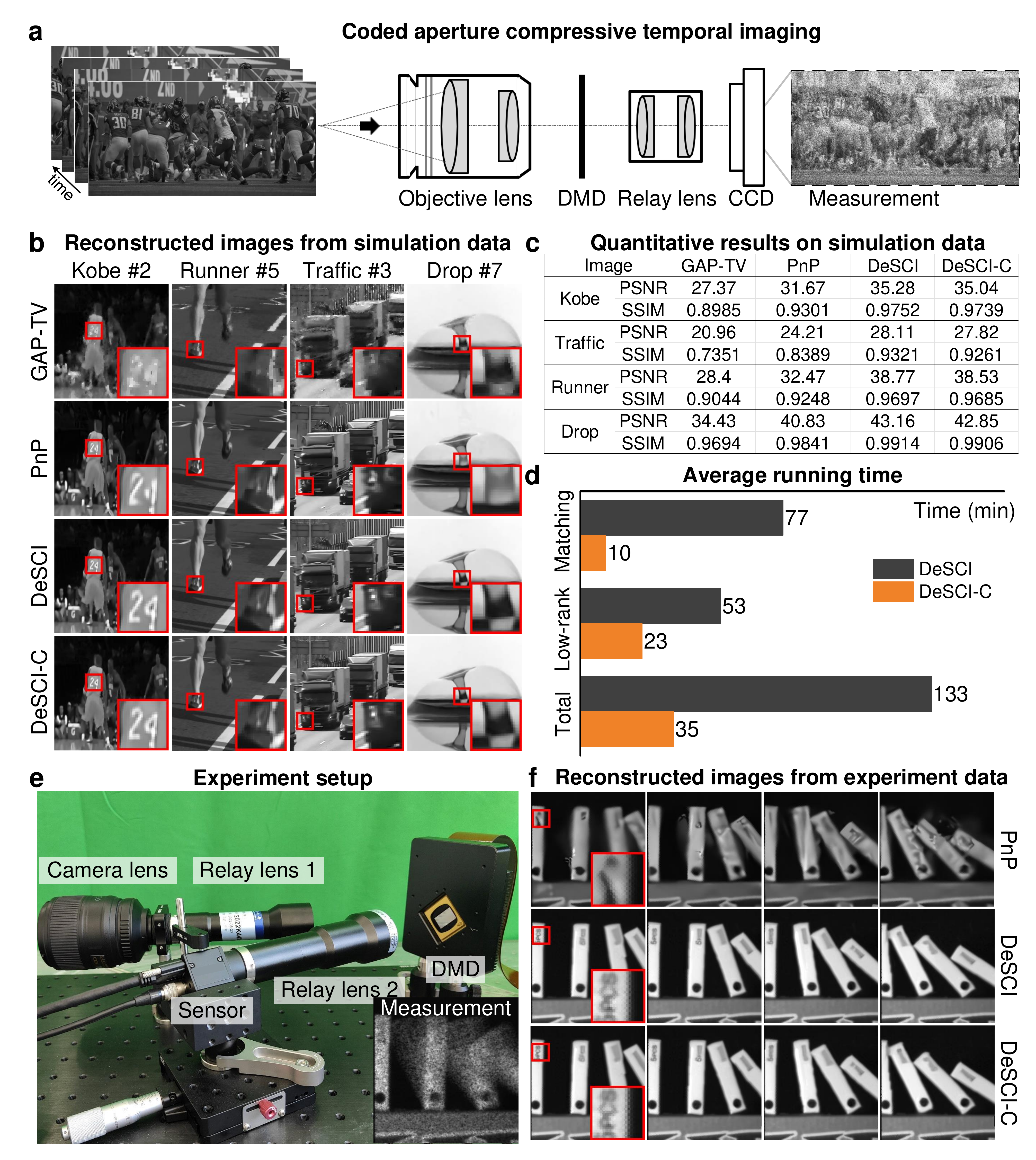}
    \caption{Reconstruction results for CACTI simulations and experiments. \textbf{a}, The imaging scheme of CACTI. \textbf{b}, Reconstruction results on benchmark simulation datasets (\texttt{Kobe, Runner, Traffic, Drop}). \textbf{c}, The average reconstruction accuracy on the test data of different algorithms. \textbf{d}, The average running time of nonlocal algorithms with and without corner-based exemplar patch selection. \textbf{e}, The proof-of-concept prototype of the CACTI system. \textbf{f}, The reconstruction results of the real-captured measurement via different algorithms.}
    \label{fig:sci}
\end{figure*}

CACTI\cite{llull2013coded27} is a typical temporal compressive imaging modality. As shown in Fig. 2 \textbf{a}, CACTI compresses temporally continuous 2D scenes into one snapshot via aperture coding. Mathematically, it can be modeled as
\begin{equation}
	\boldsymbol{y} = \sum_{i=1}^N{\boldsymbol{A}_{i}}{\boldsymbol{x}_{i}}
	\label{eq:2}
\end{equation}
where $\boldsymbol{{x}_{i}}$ is the scene at moment $\boldsymbol{i}$, $\boldsymbol{A}$ denotes the spatial variant mask generated by the digital micro-mirror device (DMD), and $\boldsymbol{y}$ represents the measurement.

We test the corner-based exemplar patch selection in the state-of-the-art NLR-based algorithm DeSCI\cite{liu2018rank20} for CACTI reconstruction. Denote DeSCI with corner-based BM as DeSCI-C. Accept from the corner-based exemplar patches, a modest number of uniformly selected patches are included. The interval of two neighbor exemplar patches is 3 times larger than vanilla BM. The common-used benchmark datasets (\texttt{Kobe, Runner, Traffic, Drop})\cite{yuan2020plug18,ma2019deep30} are applied for simulation. One snapshot of the simulation data encodes 8 frames of the scenes at 256 × 256 resolution. The commonly used CS-based algorithms GAP-TV\cite{yuan2016generalized25} and Plug-and-play (PnP)\cite{yuan2020plug18} are included as the baselines. FFDNet\cite{zhang2018ffdnet31} is employed for the learning-based regularization of PnP. All the hyper-parameters of above-mentioned algorithms are set as the default values provided by ref. \cite{yuan2020plug18}. Sub-figures \ref{fig:sci} \textbf{b} \& \textbf{c} show the visual and quantitative results of the simulation data. It can be seen that DeSCI with corner-based BM achieved competitive results compared to conventional DeSCI. The differences between the average PSNR and SSIM of DeSCI and DeSCI-C are 0.27 dB and 0.002, respectively. This small numerical difference is hard to be noticed in visual perception as shown in Fig. \ref{fig:sci} \textbf{b}. Furthermore, we build a proof-of-concept prototype shown in Fig. \ref{fig:sci} \textbf{e}. The prototype consists of one camera lens, two relay lenses, one DMD to generate the spatial variant masks, and one CCD sensor to capture the measurement. With the real-captured measurement encoded with 10 masks, we retrieved 10 temporal successive frames of the target scene. It can be seen from Fig. \ref{fig:sci} \textbf{f} that DeSCI and DeSCI-C can recover more sharp textures compared to the prevalent PnP method. DeSCI-C obtained similar results as DeSCI, which proved the effectiveness of corner-based BM. However, DeSCI-C has higher reconstruction efficiency due to fewer exemplar patches. The matching speed of corner-based BM is 7.7 times faster compared to vanilla BM, and the total reconstruction speed of DeSCI-C is 3.8 times faster than that of DeSCI, according to the average running times listed in Fig. \ref{fig:sci} \textbf{d}. All the above simulations and experiments demonstrate the efficiency of structural feature attention from corner-based exemplar patch selection.

\begin{figure}[b]
    \centering
    \includegraphics[width=1\linewidth]{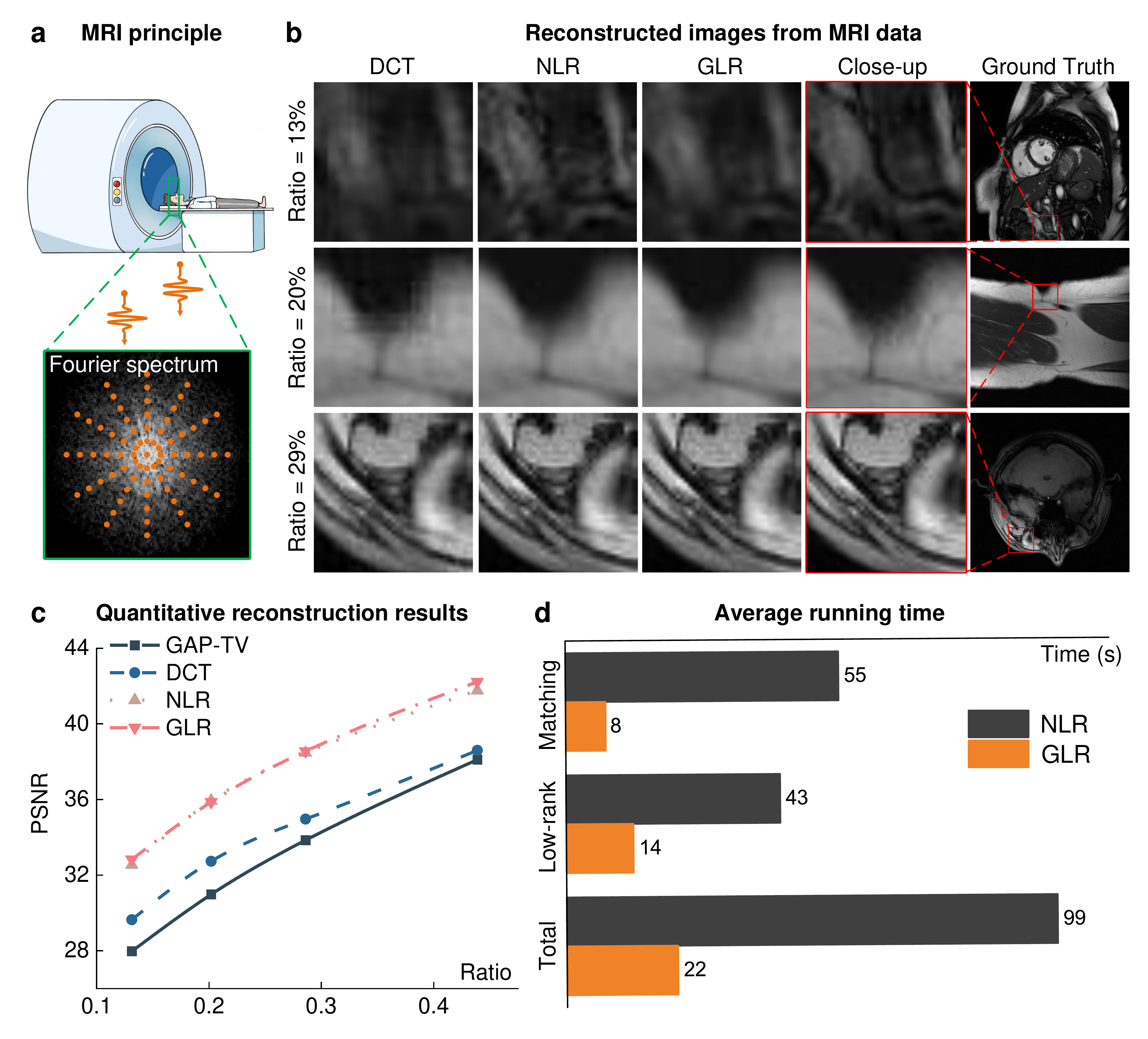}
    \caption{MRI reconstruction results. \textbf{a}, The mathematical principle of MRI. We focus on the radial sampling strategy in the Fourier domain. \textbf{b}, The visual comparison of reconstructed images (256×256 pixels) from Dong et al.’s\cite{dong2014compressive16} dataset and fastMRI\cite{zbontar2018fastmri33,knoll2020fastmri34}. \textbf{c}, The quantitative results of different reconstruction techniques. \textbf{d}, Average running time for each procedure of NLR and GLR.}
    \label{fig:mri}
\end{figure}

\subsection{Magnetic resonance imaging}
The Fourier spectrum of natural images has the nature of sparsity. One can reconstruct target scenes with magnetic resonance data even below the Nyquist sampling ratio via CS algorithms\cite{lustig2008compressed32}. As shown in Fig. \ref{fig:mri} \textbf{a}, the Fourier domain compressive imaging model is formulated as
\begin{equation}
	\boldsymbol{y} = \boldsymbol{M}\mathscr{F}(\boldsymbol{x})
	\label{eq:3}
\end{equation}
where $\boldsymbol{x}$ denotes the target scene, $\mathscr{F}$ is the Fourier transform function, $\boldsymbol{M}$ represents the sampling mask, and $\boldsymbol{y}$ is the corresponding measurement. We evaluated the reported reconstruction technique on magnetic resonance imaging (MRI). In the experiments, the sampling strategy is set as the prevalent radial mask. We applied Dong et al.’s NLR algorithm\cite{dong2014compressive16} for MRI reconstruction. GLR is formed by replacing BM with GM in NLR.

Figure \ref{fig:mri} shows the experiment results. GLR holds a similar performance to NLR in reconstruction accuracy (Fig. \ref{fig:mri} \textbf{b}, \textbf{c}). For single-channel reconstructions, GM exhibits about 7 times acceleration against BM. The low-rank approximation of GLR shows a 3 times speed-up compared to the conventional approach. GLR has an average 4.5 times improvement in running time compared to NLR. Specifically, the speed-up of low-rank approximation mainly comes from the reduction of exemplar patches by corner-based exemplar patch selection. The acceleration of the matching process results from the joint promotion of both corner-based exemplar patch selection and convolution-based structural matching. 


\begin{figure}[h!]
	\centering
	\includegraphics[width=0.95\linewidth]{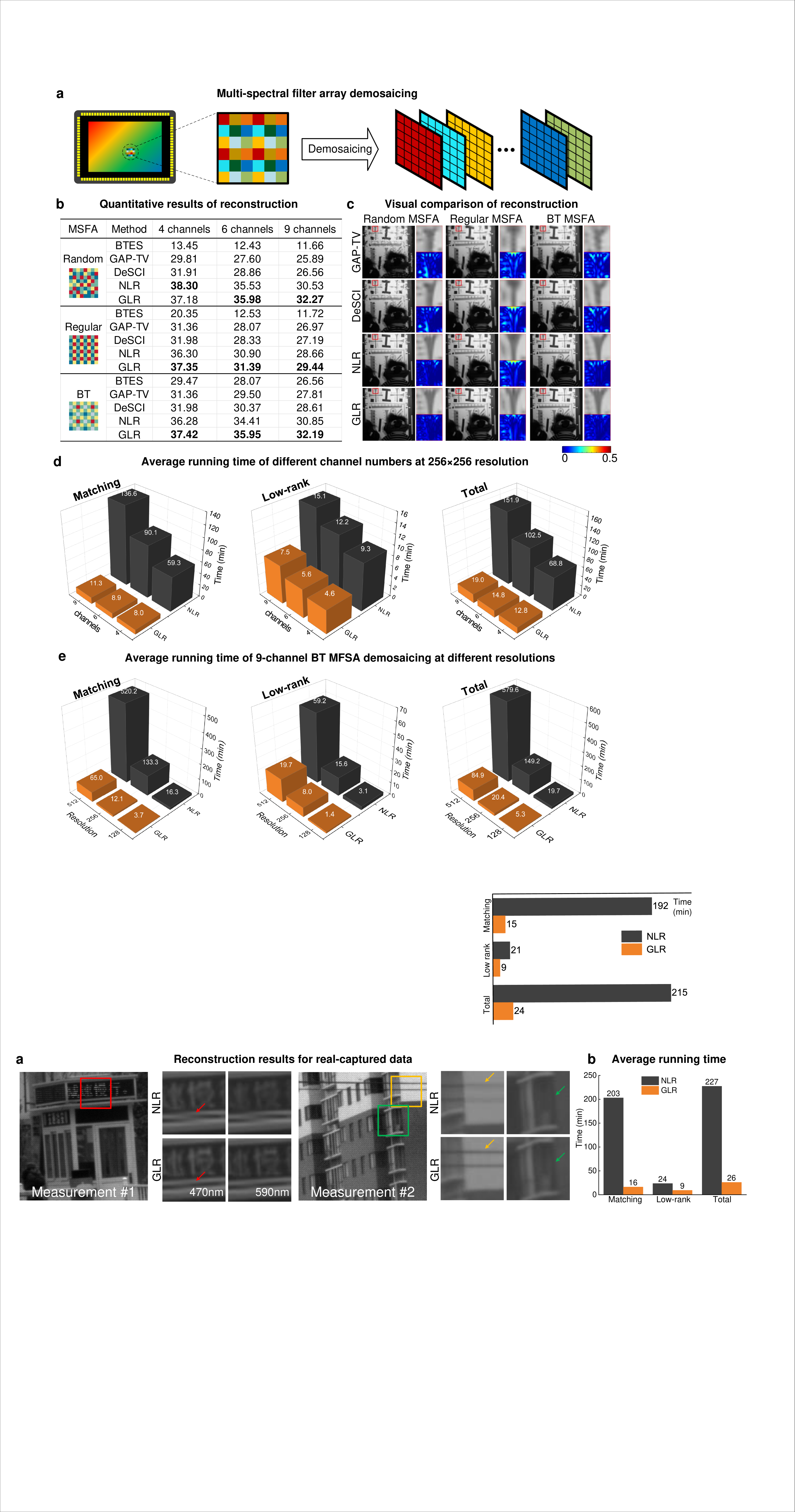}
	\caption{The demosaicing simulations under different MSFAs. \textbf{a}, The demosaicing process for a multispectral mosaic camera. \textbf{b}, The PSNR values of reconstructed multi-channel images. \textbf{c}, The reconstruction images for various 6-channel MSFAs. The closeups and corresponding error maps are shown on the right side of each image. \textbf{d}, Average running time (200 iterations in total) of different channels at 256 × 256 resolution for each procedure of NLR and GLR. \textbf{e}, Average running time of 9-channel BT MSFA demosaicing at different resolutions. }
	\label{fig:msfa_simu}
\end{figure}

\begin{figure}[h!]
	\centering
	\includegraphics[width=1\linewidth]{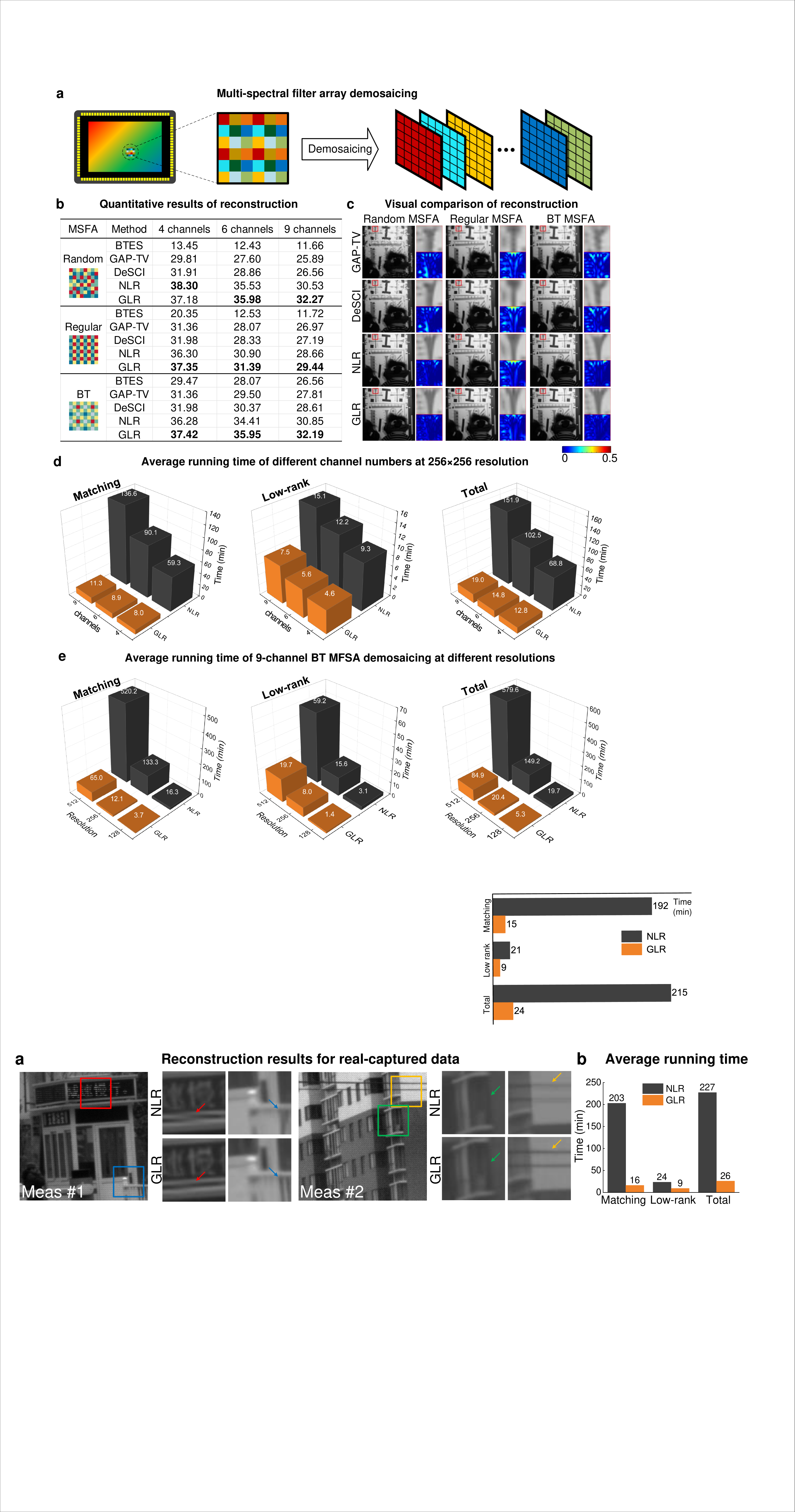}
	\caption{The demosaicing results for real-captured data. The measurement captured with a 4$\times$4 IMEC snapshot camera is provided by Feng et al. \cite{feng2021mosaic41}. The measurements are of 256$\times$256$\times$16 pixels. \textbf{a}, Visual comparisons between reconstruction results of NLR and GLR. \textbf{b}, Average running time of both algorithms.}
	\label{fig:msfa_real}
\end{figure}

\begin{figure}[h!]
	\centering
	\includegraphics[width=1\linewidth]{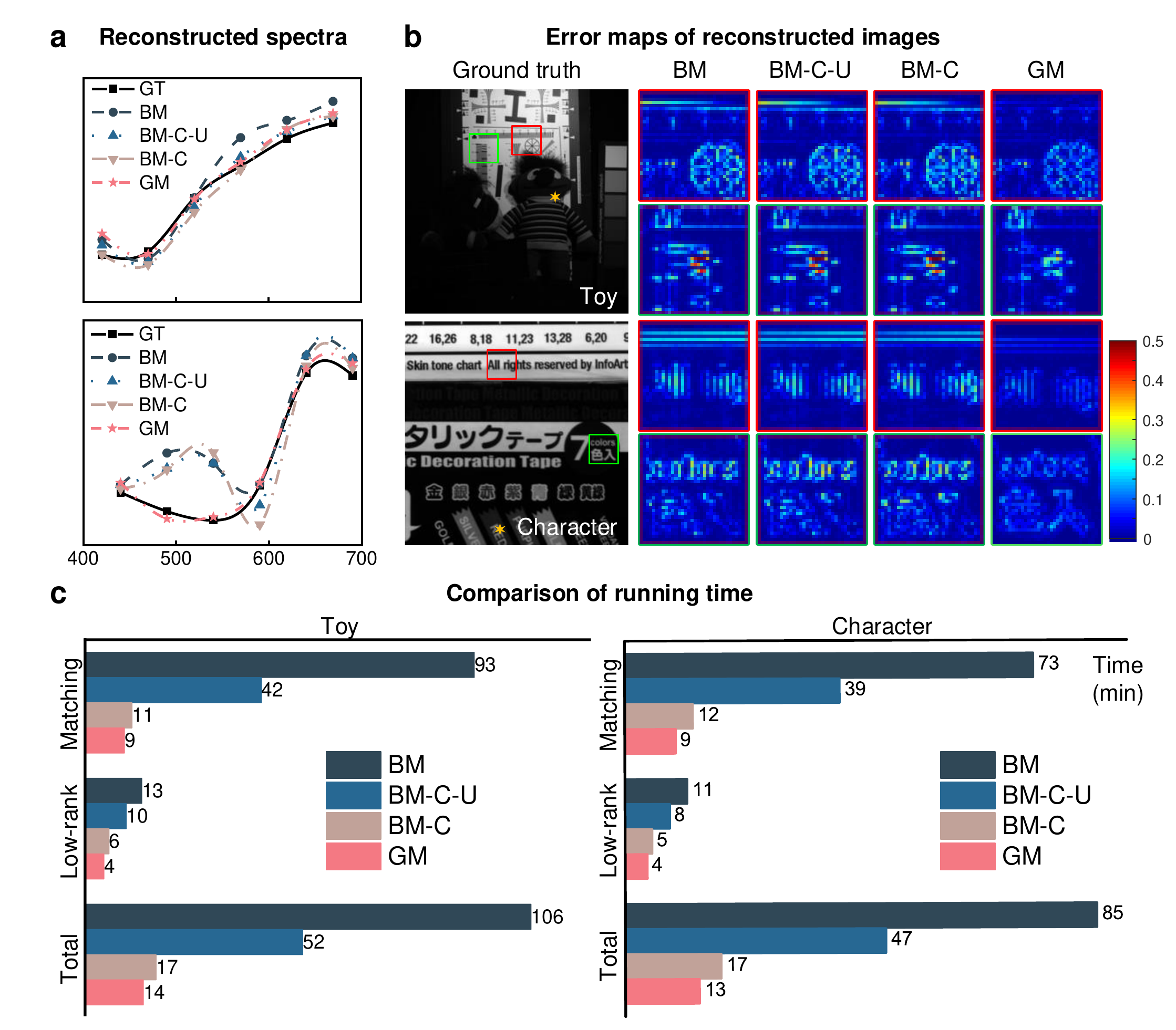}
	\caption{Comparison of various matching strategies. \textbf{a}, The spectra of the randomly selected point (marked with a hexagon in the ground truth image of \textbf{b}) in the reconstructed multispectral scene. \textbf{b}, Error maps of reconstructed images with different matching strategies "BM-C" and "BM-C-U" stand for block matching with corner-based exemplar patches and with both corner-based exemplar patches and the same number of uniform ones. \textbf{c}, Comparison of running time of nonlocal techniques with different matching strategies.}
	\label{fig:discussion1}
\end{figure}

\subsection{Multi-spectral filter array demosaicing}
Multi-spectral filter array (MSFA) is an extension of the conventional color camera that covers a Bayer filter\cite{sharma2017digital35}. MSFA cameras can acquire multi-spectral information in one snapshot\cite{cao2016computational36}. Compared to other existing multi-spectral imaging systems\cite{huang2022spectral37}, it provides an integrated, low-cost, exact registration, and full frame rate solution\cite{miao2006binary38,monno2014multispectral39}. The imaging model is denoted as
\begin{equation}
	\boldsymbol{y} = \sum_{i=1}^N{\boldsymbol{A}_{i}}{\boldsymbol{x}_{i}}
	\label{eq:4}
\end{equation}
where $\boldsymbol{A}_{i}$ stands for the filter mask of the $i$-th spectral channel. The masks are designed to be orthogonal,
\begin{equation}
	{\boldsymbol{A}_{i}}{\odot}{\boldsymbol{A}_{j}}=0
	\label{eq:5}
\end{equation}
where $\odot$ denotes the Hadamard product. However, there exhibits a fundamental compromise between spatial and spectral resolution for MSFA camera. There is a rising challenge to reconstruct full-resolution images from multi-spectral mosaic data because it is more ill-posed than demosaicing for RGB cameras (Fig. \ref{fig:msfa_simu} \textbf{a}). Bian et al.\cite{bian2021generalized40} experimentally proved that NLR algorithms with multi-channel block matching can achieve state-of-the-art performance for multi-spectral demosaicing. Inheriting Bian et al.’s framework, GLR replaces the matching strategy with multi-channel global matching (see Supplement 1). Here we conduct a series of simulations and experiments to demonstrate the superiority of GLR over conventional NLR.

As shown in both simulations and experiments of Fig. \ref{fig:msfa_simu} \textbf{b}, \textbf{c}, and \textbf{e}, the visual and numerical results indicate GLR outperforms NLR with various channel numbers and types of MSFA. The recovered closeups in Fig. \ref{fig:msfa_simu} \textbf{b} show that global matching contributes to the recovery of more delicate textures and details than conventional block matching. In multi-channel cases, GM presents obvious efficiency superiority over vanilla BM. GM provides more speed boost as the channel number increases (Fig. \ref{fig:msfa_simu} \textbf{d}). It is mainly attributed to the high efficiency of batch convolution. For large-scale reconstruction with more than 256×256 spatial resolution and 6 channels, GM achieves up to above one magnitude of improvement in running time (Fig. \ref{fig:msfa_simu} \textbf{d}, \textbf{e}). Besides, with corner-based exemplar patch selection, GM extracts fewer groups of self-repetitive patches than BM. Corner-based exemplar patch selection enables decreasing the running time of both matching and low-rank approximation. Combining the above two improvements, GLR achieves about 4 to 8 times efficiency gain while obtaining state-of-the-art performance. Taking the single-channel cases of MRI into consideration, GLR exhibits an increasing priority over NLR in terms of both running time and reconstruction accuracy as the channel number goes larger. Experiments of real-captured data validated GLR’s advancement in both accuracy and efficiency (Fig. \ref{fig:msfa_real}).

\section{Conclusion and discussion}
In this work, we engaged in a generalized global low-rank optimization for computational reconstruction. It enables fast calculation and global self-similarity regularization, offering an impetus to efficient nonlocal techniques. As validated by extensive simulations and experiments on the above three computational imaging modalities (CACTI, MRI, and MSFA demosaicing), the core contributions of GLR (structural feature attention and neural-based matching) can promote the running efficiency of nonlocal techniques by nearly an order of magnitude. Benefiting from inherent fusion of deep learning strategies and iterative optimization, GLR exhibits unique advantages in efficient and high-fidelity reconstruction for high-dimensional large-scale computational imaging tasks.

\subsection{From block matching to global matching}
To reveal the core advantages of GLR, we evaluated GM, uniform, corner-based, and corner-based+uniform BM for 6-channel BT MSFA demosaicing. For the corner-based+uniform BM, the number of uniformly picked exemplar patches is set to be similar to that of corner-based patches. As shown in Fig. \ref{fig:discussion1}, GLR/GM outperforms other approaches in both reconstruction accuracy and efficiency. The reconstructed images with GM exhibit more sharp textures (Fig. \ref{fig:discussion1} \textbf{b}) and high-fidelity spectra (Fig. \ref{fig:discussion1} \textbf{a}) especially in the regions of rich textures. NLR with corner-based block matching has less running time than with uniform block matching, due to fewer exemplar patches. However, too few self-repetitive patches for block matching leads to reduced accuracy (Fig. \ref{fig:discussion1} \textbf{b}). With similar numbers of corner-based exemplar patches, the comparisons between GM and corner-based BM indicate that GM outperforms BM, demonstrating the significance of global structural self-similarity.

\begin{figure}[t]
	\centering
	\includegraphics[width=1\linewidth]{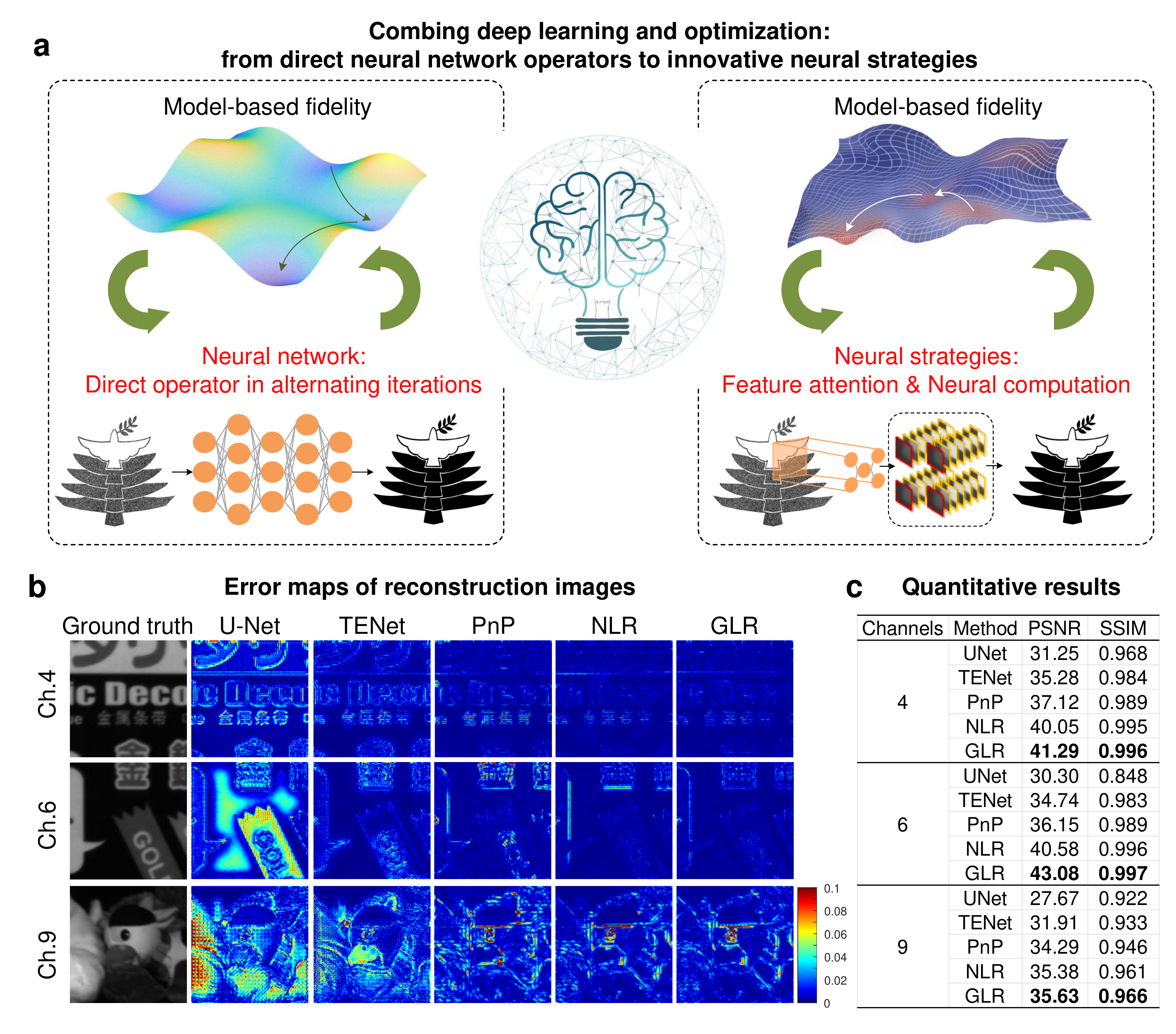}
	\caption{Comparison between end-to-end networks, PnP, and GLR. \textbf{a}, The diagrams of PnP (direct neural network operator) and GLR (neural strategies) schemes that combine deep learning and optimization. \textbf{b}, The error maps of BT MSFA demosaicing images from Monno’s dataset\cite{monno2014multispectral39}. \textbf{c}, The quantitative comparisons of different methods. We compare the reported GLR with end-to-end networks (U-Net and TENet trained on CAVE\cite{yasuma2010generalized46} and Monno’s dataset), PnP (FFDNet as the regularization network), and conventional NLR.}
	\label{fig:discussion2}
\end{figure}

\subsection{Combining deep learning and optimization}
By directly incorporating the neural network as a regularization operator into optimization, PnP\cite{yuan2020plug18,chang2022plug43} alternatives iterations between the model-based fidelity and learning-based regularization (Fig. \ref{fig:discussion2} \textbf{a}). The optimization framework improves the interpretability of neural networks. PnP techniques have achieved tremendous success in computational reconstruction. However, the enhancing process of the network in PnP iterations remains unexplainable due to the black-box nature of neural networks. Different from PnP, GLR presents the innovative neural strategies that combine feature detection and neural computation for explainable global low-rank optimization. The use of the ‘Conv2d’ neural operator here has clear physical significance. The ‘Conv2d’ between edge maps and exemplar edge patches measures the similarity between these inputs. Besides, the commonly used enhancing networks (such as FFDNet\cite{zhang2018ffdnet31}) for PnP have limited effective receptive field, while GLR holds global perception. As shown in Fig. \ref{fig:discussion2} \textbf{b}, GLR outperforms PnP, end-to-end networks (U-Net\cite{ronneberger2015u44} and TENet\cite{qian2019trinity45}), and conventional NLR in reconstruction accuracy. Above all, GLR provides a generalized and effective perspective of explainable applications of deep learning strategies.

\subsection{Outlook}
GLR can be extended to various challenging reconstruction tasks, especially in high-dimensional and high-precision applications. To further improve the calculating speed, we can run the technique on convolution-accelerate hardware such as FPGA\cite{rahman2016efficient47,zhang2018caffeine48} and specific SOC\cite{hegde2017caffepresso49,meloni2018neuraghe50}. Next, the low-complexity low-rank approximation deserves further study under the framework. The low-rank network may be a promising alternative for further promoting efficiency. What’s more, the collision between deep learning and classical optimization will lead to a spark of inspiration. The bloom of learning-based methods can drive the progress of optimization techniques.

\bibliography{sample}

\end{document}